\documentclass[aps,prl,twocolumn,showpacs,superscriptaddress,groupedaddress]{revtex4}

\usepackage{amsthm}
\usepackage{graphicx}
\usepackage{amsfonts}
\usepackage{amssymb}
\usepackage{times,txfonts}
\usepackage{hyperref}
\usepackage{color}

\begin{document}

\title{Role of non-Markovianity and backflow of information in the speed of quantum evolution}

\author{Marco Cianciaruso}
\email{cianciaruso.marco@gmail.com}
\affiliation{Centre for the Mathematics and Theoretical Physics of Quantum Non-Equilibrium Systems,
\mbox{School of Mathematical Sciences, The University of Nottingham, University Park,
Nottingham NG7 2RD, United Kingdom}}

\author{Sabrina Maniscalco}
\email{smanis@utu.fi}
\affiliation{\mbox{Turku Centre for Quantum Physics, Department of Physics and Astronomy,
University of Turku, FI-20014, Turun Yliopisto, Finland}}
\affiliation{\mbox{Centre for Quantum Engineering, Department of Applied Physics, Aalto University, P.O. Box 11000, FIN-00076 Aalto, Finland}}

\author{Gerardo Adesso}
\email{gerardo.adesso@nottingham.ac.uk}
\affiliation{Centre for the Mathematics and Theoretical Physics of Quantum Non-Equilibrium Systems,
\mbox{School of Mathematical Sciences, The University of Nottingham, University Park,
Nottingham NG7 2RD, United Kingdom}}

\begin{abstract}
We consider a two-level  open quantum system undergoing either pure dephasing, dissipative, or multiply decohering dynamics and show that, whenever the dynamics is non-Markovian, the initial speed of evolution is a monotonic function of the relevant physical parameter driving the transition between the Markovian and non-Markovian behaviour of the dynamics. In particular, within the considered models, a speed increase can only be observed in the presence of backflow of information from the environment to the system. %
%the speed can increase %We further show that, in the considered models, the speed can
\end{abstract}

\date{April 25, 2017}
\maketitle

\noindent\emph{\bfseries Introduction.} The inevitable interaction between any system and its surroundings makes the study of open system dynamics indispensable~\cite{davies1976theory,alicki1987quantum,breuer2002theory}. This is especially true within the quantum realm, wherein the environment has in general a major detrimental effect on the quantum features of the system and thus hinders the performance of quantum technologies~\cite{zurek2003decoherence,schlosshauer2007decoherence}. However, when the system-environment correlation time approaches any of the time-scales characterizing the system dynamics, i.e., within the so-called non-Markovian regime~\cite{wolf2008assessing}, it may happen that reservoir memory effects give rise to revivals of the quantum properties of the system, a phenomenon that is known as backflow of information from the environment to the system~\cite{breuer2009measure,rivas2010entanglement,lu2010quantum,vasile2011quantifying,chruscinski2011measures,luo2012quantifying,lorenzo2013geometrical,bylicka2014non,dhar2015characterizing,souza2015gaussian}. The possibility to exploit the environment itself to combat decoherence is one of the many reasons that have recently attracted a tremendous interest into the characterization, detection, and quantification of non-Markovian dynamics~\cite{breuer2012foundations,rivas2014quantum,breuer2016colloquium,devega2017dynamics}.

In particular, some effort has been recently devoted to investigating the role played by non-Markovianity in the speed of evolution of a quantum system~\cite{deffner2013quantum,zhen2014quantum,sun2015quantum,meng2015minimal,mirkin2016quantum,zhang2016control}, whose control is an essential ingredient in many operational tasks \cite{frey2016quantum}. For example, when the open quantum system is used as a quantum memory, one needs longer coherence time and thus slowing down the noisy dynamics can be beneficial~\cite{julsgaard2004experimental}. On the other hand, if one is performing a quantum logic gate on the system, it is instead the speeding up of the evolution that will be desirable in order to reach the fastest possible computation time~\cite{lloyd2000ultimate}. The authors of refs.~\cite{deffner2013quantum,zhen2014quantum,sun2015quantum,meng2015minimal,mirkin2016quantum,zhang2016control,cimmarusti2015environment} investigated the effect of non-Markovianity on some instances of quantum speed limits holding for open quantum processes, expressed as lower bounds to the evolution time necessary to go from an initial state to a target state through a given noisy dynamics \cite{taddei2013quantum,del2013quantum,deffner2013quantum,pires2016generalized}. More specifically, they analyzed the tightness of these lower bounds, compared with an actual fixed evolution time, when changing the relevant physical parameter that determines the transition between the Markovian and non-Markovian regime of the dynamics. In~\cite{deffner2013quantum} it was shown that some examples of quantum speed limits can get less tight when increasing the degree of non-Markovianity of the dynamics of a two-level atom on resonance with a lossy cavity. On the other hand, in \cite{zhen2014quantum} it was shown that the same quantum speed limits adopted in~\cite{deffner2013quantum} become tighter when increasing the degree of non-Markovianity of the dynamics of the polarization degree of a photon undergoing pure dephasing due to the interaction with the frequency degrees of freedom of the photon itself. However, it is still not clear whether the fact that a quantum speed limit becomes, e.g., less tight by increasing the degree of non-Markovianity implies that also the corresponding actual evolution time is decreasing and thus non-Markovianity is speeding up the evolution.

In this paper we analyze the behaviour of the actual speed of the evolution of a two-level quantum system undergoing paradigmatic examples of  purely dephasing, dissipative and multiply decohering dynamics amenable to an analytical solution. We show that, when the dynamics is non-Markovian, the initial value of the speed of evolution is a monotonic function of the relevant physical parameter driving the transition between Markovianity and non-Markovianity of the evolution (see Table \ref{SummaryTable}), while this need not be the case when the dynamics is Markovian. More specifically, within the aforementioned models, we show that a speed-up of the evolution can only happen in the presence of information backflow from the environment to the system. This clarifies the role of specific non-Markovian signatures in achieving speed-ups, which may have relevant implications for quantum technologies.

%However, being in general the degree of non-Markovianity a non-monotonic function of the parameter driving the Markovian to non-Markovian transition, we also provide physical instances where there is no monotonic relationship between the speed of evolution and the degree of non-Markovianity.

 \begin{table*}[bt]
 \centering
\begin{tabular}{ccccccc}
\hline \hline
Dynamics & CP-indivisible & Backflow of information  & \multicolumn{2}{c}{$v^2(0)$} &  &   \\ \hline
Ohmic Pure Dephasing & $s > 2$ & $s > 2$ &  $2 \omega_c^2 \Gamma[s+1] \sin^2\theta$ &  &   \\
Photon Polarization Pure Dephasing & $\xi_1 \leq \xi \leq \xi_2$ \cite{zhen2014quantum}  & $\xi_1 \leq \xi \leq \xi_2$ \cite{zhen2014quantum} & $\frac{1}{2} (\Delta n)^2 \left[2 \sigma^2 + \omega_1^2 + \omega_2^2 - \left(\omega_2^2 - \omega_1^2\right) \cos2\xi\right] \sin^2\theta$ &  &  \\
Jaynes-Cummings on resonance & $\gamma_M > \lambda/2$ & $\gamma_M > \lambda/2$ &  $\gamma_M^2 \lambda^2 \sin^4\frac{\theta}{2}$ &  &  \\
Jaynes-Cummings with detuning & \cite{li2010non} & \cite{li2010non} &  $(4W^2 - \Delta^2) \sin^4\frac{\theta}{2}$ &  &  \\
Pauli Channel with $\gamma_3(t)=-(\omega/2)\tanh(\omega t)$ & $0 < \omega \leq \lambda$ & - & $-4 \lambda^2 \cos^2\theta - (\lambda^2+\omega^2)\sin^2\theta$ &  &  \\
Pauli Channel with $\gamma_3(t)=(\omega/2)\tan(\omega t)$ & $\omega > 0$ & $\omega > 0$ & $-4 \lambda^2 \cos^2\theta - (\lambda^2-\omega^2)\sin^2\theta$ &   &  \\
\hline \hline
\end{tabular}

\caption{\label{SummaryTable} A summary of the regions of CP-indivisibility and presence of backflow of information (as defined in \cite{breuer2009measure}), as well as of the squared initial speed of evolution corresponding to paradigmatic instances of single-qubit purely dephasing, dissipative and multiply decohereing dynamics as a function of the relevant physical parameter driving the transition between Markovian and non-Markovian regime.}
\end{table*}

\smallskip
\noindent\emph{\bfseries Non-Markovian dynamics.} The evolution $\rho(t)$ over the time interval $t$ of any initial quantum state $\rho(0)$  can be characterized by a one-parameter family $\{\Lambda_t | t \geq 0, \Lambda_0 = \mathbb{I}\}$ of completely positive and trace preserving (CPTP) maps, so-called dynamical maps, as follows~\cite{nielsen2010quantum}:
\begin{equation}\label{eq:DynamicalMap}
\rho(t) = \Lambda_t [\rho(0)].
\end{equation}
If the quantum system is closed, it undergoes a reversible unitary evolution so that the corresponding family of dynamical maps is a one-parameter group, i.e.: (i) it contains the identity element; (ii) it is closed under composition of any two elements; (iii) such composition is associative; (iv) the inverse $\Lambda_t^{-1}$ of every element exists and is also an element of the family. On the other hand, when the quantum system is open, it undergoes a noisy irreversible evolution, which prevents the corresponding family of dynamical maps from being a group, as  property (iv) is inevitably violated, i.e., either $\Lambda_t^{-1}$ does not exist for some $t$ or if $\Lambda_t^{-1}$ does exist for any $t$ it is not contained within the family of dynamical maps describing the evolution.

A particular and well-known class of open evolutions is such that the corresponding family of dynamical maps forms a one-parameter semi-group~\cite{gorini1976completely,lindblad1976generators}, which satisfies all the remaining properties (i), (ii) and (iii). The semi-group property of a family of dynamical maps can be succinctly characterized by the following relation:
%\begin{equation}\label{eq:SemiGroupProperty}
$\Lambda_{t} = \Lambda_{t-s}\Lambda_s$,
%\end{equation}
holding for any $0 \leq s \leq t$. This property means that the map can be divided into infinitely many identical steps, in such a way that the ensuing dynamics can be intuitively interpreted as being memoryless. This class of open evolutions represents the prototypical example of Markovian dynamics.

The semi-group property can be easily generalized by introducing the notion of CP-divisibility~\cite{rivas2010entanglement,chruscinski2014degree,torre2015non,liuzzo2016non}. The dynamics $\{\Lambda_{t}\}$ is said to be CP-divisible if there exists a two-parameter family $\{\tilde{\Lambda}_{t,s}\}$ of CPTP maps, which need not be within the family $\{\Lambda_{t}\}$, such that:
\begin{equation}\label{eq:CPDivisibilityProperty}
\Lambda_{t} = \tilde{\Lambda}_{t,s}\Lambda_s,
\end{equation}
for any $0\leq s \leq t$. Analogously to the case where the family of dynamical maps forms a semi-group, a CP-divisible dynamics can be seen as the concatenation of infinitely many other dynamical maps and thus can be loosely interpreted as being memoryless and is commonly considered to be Markovian.

Yet the border between Markovian and non-Markovian dynamics is still elusive as consensus on where to draw it, and on how to quantify non-Markovianity of maps beyond such border, is still lacking in the current literature~\cite{breuer2012foundations,rivas2014quantum,breuer2016colloquium,devega2017dynamics}. On one hand, in the CP-divisibility paradigm one may quantify the non-Markovianity degree by measuring how much the intermediate map $\tilde{\Lambda}_{t,s}$ appearing in Eq.~(\ref{eq:CPDivisibilityProperty}) is far from being CPTP~\cite{rivas2010entanglement,chruscinski2014degree,torre2015non,liuzzo2016non}. On the other hand, one may consider the backflow of information from the environment to the system as a genuinely non-Markovian signature. Information manifests itself in many forms, such as quantum state distinguishability, coherence, and correlations. All these manifestations of information share a common property, i.e., being contractive under CPTP maps, which is due to the fact that CPTP maps are the mathematical counterpart of noise and thus can only produce a loss of information. However, if the dynamics is not CP-divisibile, the fact that the intermediate map $\tilde{\Lambda}_{t,s}$ is not CPTP may give rise to temporary revivals of information throughout the evolution. This alternative paradigm thus estimates the degree of non-Markovianity by measuring how much information flows back to the system during the entire evolution~\cite{breuer2009measure,rivas2010entanglement,lu2010quantum,vasile2011quantifying,chruscinski2011measures,luo2012quantifying,lorenzo2013geometrical,bylicka2014non,dhar2015characterizing,souza2015gaussian}.
When considering this paradigm in the following, we will specifically adopt the indicator of backflow of information based on trace distance as introduced in \cite{breuer2009measure}.

\smallskip
\noindent\emph{\bfseries Speed of quantum evolution.} Information theory stands as the fundamental bridge linking non-Markovianity of a dynamics with the speed of the corresponding evolution~\cite{anandan1990geometry,taddei2013quantum,toth2014quantum,pires2016generalized}. The latter can be indeed naturally introduced by resorting to any CPTP-contractive Riemannian metric $\textbf{g}$ defined on the set of quantum states, which assigns to the neighbouring states $\rho$ and $\rho+d\rho$ the squared infinitesimal distance
\begin{equation}\label{eq:InfinitesimalDistance}
(ds)^2 = \textbf{g}_\rho(d\rho,d\rho).
\end{equation}
Indeed, by using Eq.~(\ref{eq:InfinitesimalDistance}), the speed of the quantum evolution $\rho(t)=\Lambda_t[\rho(0)]$ at time $t$ can be immediately defined as
\begin{equation}\label{eq:SpeedofEvolution}
v(t) = \frac{ds}{dt} = \sqrt{g(t)},
\end{equation}
where $g(t)=\textbf{g}_{\rho(t)}(\dot{\rho}(t),\dot{\rho}(t))$. The Morozova-Chencov-Petz theorem states that there are infinitely many such metrics~\cite{petz1996monotone}, two paradigmatic examples of which being the Bures-Uhlmann metric~\cite{uhlmann1993density,uhlmann1995geometric}, also known as quantum Fisher information metric, and the Wigner-Yanase metric~\cite{gibilisco2003wigner}. In this paper we will adopt the former, for which the following useful relation holds as well~\cite{braunstein1994statistical}:
\begin{equation}\label{eq:QFIintermsofFidelity}
g(t)=- 2 \frac{d^2}{dt^2} F(\rho(0),\rho(t)),
\end{equation}
where $F(\rho,\sigma)=\left(\mbox{Tr}(\sqrt{\rho}\sigma\sqrt{\rho}) \right)^2$ is the Uhlmann fidelity between the states $\rho$ and $\sigma$.

We now investigate the behaviour of the initial speed of evolution of a two-level quantum system undergoing typical dynamics.
We will impose the initial condition $\rho(0)=|\psi\rangle\langle\psi|$, corresponding to the qubit being in an arbitrary pure state
$|\psi\rangle=\cos\frac{\theta}{2} |0\rangle + e^{i\phi}\sin\frac{\theta}{2} |1\rangle $,
with Bloch vector
\begin{equation}\label{eq:BlochVectorInitialState}
\vec{n}(0)=\{\sin\theta\cos\phi,-\sin\theta\sin\phi,\cos\theta\},
\end{equation}
where $\theta\in[0,\pi]$ and $\phi\in[0,2 \pi[$.
We will calculate the fidelity between $\rho(0)$ and $\rho(t)$ via the general formula~\cite{hubner1992explicit,jozsa1994fidelity}
\begin{eqnarray}\label{eq:FidelityGeneralCase}
\!\!\!\!\!&&\!\!\!F(\rho(0),\rho(t)) \\
\!\!\!\!\!&&= \frac{1}{2} \left[1 + \vec{n}(0)\cdot\vec{n}(t) + \sqrt{(1 - \vec{n}(0)\cdot\vec{n}(0)) (1 - \vec{n}(t)\cdot\vec{n}(t))}\right], \nonumber
\end{eqnarray}
where $\vec{n}(t)$ is the Bloch vector of the evolved state $\rho(t)$.

\smallskip
\noindent \emph{\bfseries Results for purely dephasing dynamics.} We begin by considering a purely dephasing dynamics, described by the following time-local master equation~\cite{breuer2002theory}:
\begin{equation}\label{eq:MasterEquationPureDephasing}
\dot{\rho}(t) = \gamma(t) (\sigma_z \rho(t) \sigma_z - \rho(t)),
\end{equation}
where $\gamma(t)=-\dot{G}(t)/G(t)$ is the decay rate and $G(t)$ is the decoherence function accounting for all the environmental features relevant to the system dynamics. Inserting into Eq.~(\ref{eq:FidelityGeneralCase}) the Bloch vector of the initial state, given by Eq.~(\ref{eq:BlochVectorInitialState}), and of the corresponding evolved state, given by $\vec{n}(t)=\left\lbrace \mbox{Re}\left(e^{i\phi} G(t)\right) \sin\theta,-\mbox{Im}\left(e^{i\phi} G(t)\right) \sin\theta ,\cos\theta\right\rbrace$,  we get
%\begin{equation}\label{eq:FidelityPureDephasing}
$F(\rho(0),\rho(t)) = \frac{1}{4}\left[3+ \cos 2\theta + 2 \mbox{Re}(G(t))\sin^2\theta  \right]$.
%\end{equation}
Therefore, by using Eqs.~(\ref{eq:SpeedofEvolution}) and (\ref{eq:QFIintermsofFidelity}), we immediately obtain that the squared speed of evolution at time $t$ is given by
\begin{equation}\label{eq:SpeedofEvolutionPureDephasing}
v(t)^2 = - \mbox{Re}\left(\ddot{G}(t)\right) \sin^2\theta.
\end{equation}

We explore two particular physical instances governed by the master equation of Eq.~(\ref{eq:MasterEquationPureDephasing}). We first consider a qubit interacting with a bosonic reservoir at zero temperature with Ohmic spectrum~\cite{luczka1990spin,palma1996quantum,reina2002decoherence}, whose decoherence function is
\begin{equation}\label{eq:DecoherenceFunctionOhmicPureDephasing}
G(t) = e^{-\Upsilon(t)}, \quad \mbox{with \quad $\Upsilon(t)=2 \int_0^t \gamma(t')dt'$},
\end{equation}
where $\gamma(t)=\omega_c {[1+{(\omega_c t )}^2]}^{-s/2}\Gamma[s]\sin\left[s \arctan(\omega_c t )\right]$, with
$\omega_c$ the cut-off frequency, $s$ the Ohmicity parameter, and $\Gamma[x]$ the Euler function. This dynamics is CP-divisible when $s \leq 2$, while it is CP-indivisible and manifests backflow of information for any $s>2$~\cite{breuer2009measure,rivas2010entanglement,haikka2013non}. By substituting Eq.~(\ref{eq:DecoherenceFunctionOhmicPureDephasing}) into Eq.~(\ref{eq:SpeedofEvolutionPureDephasing}), we get that the squared initial speed of evolution is
\begin{equation}
v(0)^2= 2 \omega_c^2 \Gamma[s+1] \sin^2\theta,
\end{equation}
which is a strictly monotonically increasing function of $s$ for any $s>2$, i.e., in the whole non-Markovian region, while it is not a monotonic function of $s$ anymore when $s \leq 2$.
%, i.e., in the Markovian region.

We then consider another physical example of purely dephasing dynamics, wherein the two-level open quantum system is implemented by the polarization degree of freedom of a photon with its frequency degrees of freedom playing the role of the environment, which is coupled to the system via a birefringent material~\cite{liu2011experimental,zhen2014quantum}. The corresponding decoherence function is given now by
\begin{equation}\label{eq:DecoherenceFunctionPolarizationPureDephasing}
G(t)= e^{-\frac{\sigma^2 (\Delta n)^2 t^2}{2}}(e^{i\omega_1 \Delta n t}\cos^2\xi + e^{i\omega_2 \Delta n t}\sin^2\xi),
\end{equation}
where $\Delta n$ is the difference between the birefringent material refraction indexes for a photon in the vertical and horizontal polarization, respectively, while $\sigma$, $\omega_1$, $\omega_2$ and $\xi$ are the parameters characterizing the bimodal distribution representing the probability of finding the photon in a mode with a given frequency. More specifically, $\sigma$ is the common width of the two peaks, which are centred at the frequencies $\omega_1$ and $\omega_2$, and $\xi\in[0,\pi/2]$ is the parameter controlling the relative weight of the two peaks. This dynamics is not CP-divisible and manifests backflow of information when $\xi\in[\xi_1,\xi_2]$, with $\xi_1$ and $\xi_2$ provided in~\cite{zhen2014quantum}. By plugging Eq.~(\ref{eq:DecoherenceFunctionPolarizationPureDephasing}) into Eq.~(\ref{eq:SpeedofEvolutionPureDephasing}), we get that the squared initial speed of evolution is given by
\begin{equation}
v(0)^2= \frac{1}{2} (\Delta n)^2 \left[2 \sigma^2 + \omega_1^2 + \omega_2^2 - \left(\omega_2^2 - \omega_1^2\right) \cos2\xi\right] \sin^2\theta,
\end{equation}
which is a strictly monotonically increasing (resp.~decreasing) function of $\xi$ for any $\xi\in[0,\pi/2]$ when $\omega_2>\omega_1$ ($\omega_1>\omega_2$).

\smallskip
\noindent \emph{\bfseries Results for dissipative dynamics.} Let us now study the initial speed of evolution of a qubit undergoing amplitude damping, a paradigmatic example of dissipative evolution. This is described by the following master equation~\cite{breuer2002theory}:
\begin{equation}
\dot{\rho}(t) = \gamma(t) \left(\sigma_-\rho(t)\sigma_+ - \{\sigma_+\sigma_-,\rho(t)\}/2 \right),
\end{equation}
where $\gamma(t)=-2\mbox{Re}\left(\dot{G}(t)/G(t)\right)$ is the decay rate, $G(t)$ is the decoherence function, while $\sigma_\pm=\sigma_x \pm i \sigma_y$ are the raising and lowering operators of the qubit. By imposing the initial condition Eq.~(\ref{eq:BlochVectorInitialState}), we get that the Bloch vector of the evolved state is
%\begin{eqnarray}\label{eq:BlochVectorAmplitudeDamping}
$\vec{n}(t)  %\nonumber \\
=\big\lbrace \mbox{Re}(e^{-i\phi} G(t))\sin\theta,\mbox{Im}(e^{-i\phi} G(t))\sin\theta,1-|G(t)|^2 (1-\cos\theta)\big\rbrace$. %\nonumber \\
%\end{eqnarray}
The fidelity between the evolved state and the initial state, according to Eq.~(\ref{eq:FidelityGeneralCase}),  is thus given by
%\begin{eqnarray}\label{eq:FidelityAmplitudeDamping} \nonumber
$F(\rho(0),\rho(t)) = \frac{1}{2} \left(1 + \cos\theta - 2 |G(t)|^2 \cos\theta \sin^2\frac{\theta}{2} + \mbox{Re}(G(t)) \sin^2\theta\right)$.
%,
%\end{eqnarray}
By using Eqs.~(\ref{eq:SpeedofEvolution}) and (\ref{eq:QFIintermsofFidelity}), the squared speed of evolution is then
\begin{equation}\label{eq:SpeedofEvolutionAmplitudeDamping}
\mbox{$v(t)^2 = - \mbox{Re}\left(\ddot{G}(t)\right) \sin^2\theta - 2 |G(t)|^2 \gamma(t)\cos\theta\sin^2\frac{\theta}{2}$}.
\end{equation}

We now consider the Jaynes-Cummings model as a physical implementation of an amplitude-damped qubit~\cite{breuer2002theory}. This model consists of a two-level atom immersed in a lossy cavity with Lorentzian spectral density, with decoherence function
\begin{equation}\label{eq:DecoherenceFunctionAmplitudeDamping}
G(t) =  e^{-\frac{(\lambda-i\Delta) t}{2}}\left[ \cosh\left(\frac{\Omega t}{2}\right) + \frac{\lambda-i\Delta}{\Omega} \sinh\left(\frac{\Omega t}{2}\right) \right],
\end{equation}
where $\Omega=\sqrt{\lambda^2-2i\lambda\Delta -4W^2}$, $W=\gamma_M\lambda/2+\Delta^2/4$, $\lambda$ is the the width of the reservoir spectral density,  which is centred at a frequency that is detuned from the atomic frequency by the amount $\Delta$, and finally $\gamma_M$ is the effective coupling constant.

When the Jaynes-Cummings model is on resonance, i.e., $\Delta=0$, the dynamics is divisible when $\gamma_M\leq \lambda/2$, while it gives rise to both CP-indivisibility and backflow of information for any $\gamma_M> \lambda/2$. On the other hand, by increasing the detuning $\Delta$, the threshold value of $\gamma_M/\lambda$ above which the dynamics is CP-indivisible decreases~\cite{breuer2009measure,li2010non,zeng2011equivalence}.

By replacing Eq.~(\ref{eq:DecoherenceFunctionAmplitudeDamping}) into Eq.~(\ref{eq:SpeedofEvolutionAmplitudeDamping}), we get that the squared initial speed of evolution is given by
\begin{equation}
\mbox{$v(0)^2=  (4W^2 - \Delta^2)  \sin^4\frac{\theta}{2}$},
\end{equation}
which is a strictly monotonically increasing function of $\gamma_M$.

\smallskip
\noindent \emph{\bfseries Results for multiply decohering dynamics.} We finally consider an example of multiply decohering dynamics, that is, a qubit undergoing a Pauli channel \cite{vacchini2012classical,chruscinski2013non,chruscinski2014degree,megier2016non}. This evolution can be described by the following time-local master equation:
\begin{equation}
\dot{\rho}(t) = {\sum}_{j=1}^3 \gamma_j(t) (\sigma_j \rho(t) \sigma_j - \rho(t)),
\end{equation}
where the $\gamma_j(t)$'s are the decay rates. The solution of the above master equation is given by
$\rho(t)=\sum_{j=0}^3 p_j(t) \sigma_j \rho(0) \sigma_j$,
where
$p_{0,1}(t) = \frac{1}{4} (1 + \lambda_1(t) \pm \lambda_2(t) \pm \lambda_3(t))$ and
$p_{2,3}(t) = \frac{1}{4} (1 - \lambda_1(t) \pm \lambda_2(t) \mp \lambda_3(t))$
with
$\lambda_j(t) = e^{-(\Upsilon_k(t) + \Upsilon_l(t))}$ ($j \neq k \neq l \in \{1,2,3\}$),
and
$\Upsilon_j(t) = 2 \int_0^t \gamma_j(t')dt'$. By imposing again the initial condition expressed in Eq.~(\ref{eq:BlochVectorInitialState}) and assuming $\lambda_2(t)=\lambda_1(t)$, we get that the Bloch vector of the evolved state is given by
%\begin{equation}\label{eq:BlochVectorPauliChannel}
$\vec{n}(t)=\{\lambda_1(t) \cos\phi \sin\theta, -\lambda_1(t) \sin\phi \sin\theta, \lambda_3(t) \cos\theta\}$.
%\end{equation}
The fidelity between the above evolved state $\rho(t)$ and initial state $\rho(0)$ is thus simply obtained by using Eq.~(\ref{eq:FidelityGeneralCase}), yielding:
%\begin{equation}\label{eq:fidelityversustime}
$F(\rho_0,\rho(t)) = \frac{1}{4} [2 + \lambda_1(t) + \lambda_3(t) + (\lambda_3(t) - \lambda_1(t)) \cos2\theta]$.

Therefore, by using Eqs.~(\ref{eq:SpeedofEvolution}) and (\ref{eq:QFIintermsofFidelity}), we easily get that the squared speed of evolution at time $t$ is given by
\begin{equation}\label{eq:SpeedofEvolutionPauliChannels}
v(t)^2 = - \frac{1}{2}\left[\ddot{\lambda}_1(t) + \ddot{\lambda}_3(t) + \left(\ddot{\lambda}_3(t) - \ddot{\lambda}_1(t)\right) \cos2\theta\right] .
\end{equation}

Let us first consider the case with decay rates given by $\gamma_1(t)=\gamma_2(t)=\lambda/2$ and $\gamma_3(t)=-(\omega/2)\tanh(\omega t)$, i.e.,
\begin{equation}\label{eq:DecoherenceFunctionPauliChannel1}
\lambda_1(t) = \lambda_2(t)=e^{-\lambda t} \cosh(\omega t), \quad
\lambda_3(t) = e^{-2\lambda t},
\end{equation}
where $0 \leq \omega \leq \lambda$. This dynamics is CP-divisible when $\omega=0$, while it is CP-indivisible for any $0<\omega\leq \lambda$. However, there is no backflow of information for any value of $\omega$ \cite{chruscinski2014degree,bylicka2016thermodynamic}.
By replacing Eq.~(\ref{eq:DecoherenceFunctionPauliChannel1}) into Eq.~(\ref{eq:SpeedofEvolutionPauliChannels}) we get that the squared initial speed of evolution is given by
\begin{equation}
v(0)^2= -4 \lambda^2 \cos^2\theta - (\lambda^2+\omega^2)\sin^2\theta,
\end{equation}
which is a strictly monotonically decreasing function of $\omega$.

Let us now turn to considering the case of $\gamma_1(t)=\gamma_2(t)=\lambda/2$ and $\gamma_3(t)=(\omega/2)\tan(\omega t)$, i.e.,
\begin{equation}\label{eq:DecoherenceFunctionPauliChannel2}
\lambda_1(t)  =  \lambda_2(t)=e^{-\lambda t} |\cos(\omega t)|, \quad
\lambda_3(t) = e^{-2\lambda t},
\end{equation}
where $\lambda \geq0$ and $\omega \geq 0$. This dynamics is CP-divisible when $\omega=0$, while it is both CP-indivisible and manifests backflow of information  for any $\omega>0$ \cite{chruscinski2014degree,bylicka2016thermodynamic}.
By replacing Eq.~(\ref{eq:DecoherenceFunctionPauliChannel2}) into Eq.~(\ref{eq:SpeedofEvolutionPauliChannels}) we get that the squared initial speed of evolution is given in this case by
\begin{equation}
v(0)^2= -4 \lambda^2 \cos^2\theta - (\lambda^2-\omega^2)\sin^2\theta,
\end{equation}
which is a strictly monotonically increasing function of $\omega$.

\smallskip
\noindent \emph{\bfseries Conclusions.} The control of the speed of quantum evolution is an indispensable feature in several technological applications \cite{julsgaard2004experimental,lloyd2000ultimate,sorensen2016nature}. By performing an in-depth analysis, we have shown that, whenever the dynamics is non-Markovian, the initial speed of evolution of a qubit undergoing prototypical instances of purely dephasing, dissipative and multiply decohering channels, is a monotonic function of the relevant physical parameter determining the crossover between Markovianity and non-Markovianity of the evolution (see Table~\ref{SummaryTable}), which in turn may be experimentally controlled in different settings, e.g.~in quantum optics \cite{liu2011experimental,bernardes2014experimental}. More specifically, within the considered models, we have shown that a speed-up of the evolution can only be observed in the presence of information backflow from the environment to the system (as defined in \cite{breuer2009measure}).
%as witnessed by a non-monotonicity of the trace distance quantifier \cite{breuer2009measure}.
This analysis reveals that the presence of information backflow, which is a specific facet of non-Markovianity attracting increasing interest \cite{breuer2012foundations,rivas2014quantum,breuer2016colloquium,devega2017dynamics}, may play a key role as an enhancer for quantum technologies relying on fast and accurate control of open system dynamics.

Our study sheds further light on the interplay between non-divisibility and the speed of evolution of an open quantum evolution. While previous studies were more concerned with (not necessarily saturated) lower bounds to the speed of evolution and how they are affected in the non-Markovian regime \cite{deffner2013quantum,zhen2014quantum,sun2015quantum,meng2015minimal,mirkin2016quantum,zhang2016control}, this study reveals a precise connection between the actual initial speed of evolution and the relevant model parameters driving the manifestation of non-Markovianity (CP-individisility).  Yet a general criterion determining exactly when (and under which physical conditions) an increase of the parameters driving the transition to non-Markovianity amounts to speeding up rather than slowing down the evolution is still missing and certainly deserves future investigation.

\noindent \emph{\bfseries Acknowledgements.} This work is supported by the European Research Council (ERC) Starting Grant GQCOP "Genuine Quantumness in Cooperative Phenomena" (Grant No. 637352), and by the
Foundational Questions Institute (fqxi.org) Physics of the Observer Programme (Grant
No. FQXi-RFP-1601).

\bibliographystyle{apsrev}
\bibliography{nonmarkovbib}

\end{document}